\documentclass[9pt,shortpaper,twoside,web]{ieeecolor}
\usepackage{generic}
\usepackage{cite}
\usepackage{amsmath,amssymb,amsfonts}
\usepackage{algorithmic}
\usepackage{graphicx}
\usepackage{textcomp}

\usepackage{mathrsfs}
\usepackage{hyperref}
\usepackage{float}
\usepackage{color}
\usepackage{multirow}
\usepackage{lineno}
\usepackage{framed}

\newcounter{mytempeqncnt}

\newtheorem{assumption}{Assumption}
\newtheorem{problem}{Problem}
\newtheorem{theorem}{Theorem}
\newtheorem{proposition}{Proposition}
\newtheorem{remark}{Remark}
\newtheorem{lemma}{Lemma}

\newenvironment{pfof}[1]{\vspace{1ex}\noindent{\itshape Proof of #1.}\hspace{0.3em}} {\hfill $\square$ \par}

\def\BibTeX{{\rm B\kern-.05em{\sc i\kern-.025em b}\kern-.08em
    T\kern-.1667em\lower.7ex\hbox{E}\kern-.125emX}}
    
\begin{document}
\title{Data-based Transfer Stabilization in Linear Systems}
\author{Lidong Li, Claudio De Persis, Pietro Tesi, and Nima Monshizadeh
\thanks{ The work of Lidong Li was supported by the Chinese Scholarship Council (CSC) under Grant 202006120009. The work of Pietro Tesi was partially supported by the European Union under the Italian National Recovery and Resilience Plan (NRRP) of NextGenerationEU, partnership on “Telecommunications of the Future” (PE00000001 - program “RESTART”).}
\thanks{ Lidong Li, Claudio De Persis and Nima Monshizadeh are with the Engineering and Technology Institute, University of Groningen, 9747AG, The Netherlands (e-mail: l.li@rug.nl, c.de.persis@rug.nl, n.monshizadeh@rug.nl). Pietro Tesi is with DINFO, University of Florence, 50139 Florence, Italy (e-mail: pietro.tesi@unifi.it).}
}

\maketitle

\begin{abstract}
We present a novel framework for transferring the knowledge from one system (source) to design a stabilizing controller for a second system (target). Our motivation stems from the hypothesis that abundant data can be collected from the source system, whereas the data from the target system is scarce. We consider both cases where data collected from the source system is noiseless and noisy.
For each case, by leveraging the data collected from the source system and a priori knowledge on the maximum distance of the two systems,
we find a suitable, and relatively small, compact set of systems that contains the actual target system, and then provide a controller that stabilizes the compact set.
In particular, the controller can be obtained by solving a set of linear matrix inequalities (LMIs). Feasibility of those LMIs is discussed in details.  
We complement our theoretical findings by two numerical case studies of low-order and high-order systems. 
\end{abstract}

\begin{IEEEkeywords}
    Data-driven control, linear systems, transfer learning, transfer stabilization
\end{IEEEkeywords}

\section{Introduction}
Learning controllers directly from data is a fertile research direction and has witnessed a surge in interest in recent years. This provides a compelling alternative to the indirect approach of a sequential system identification and model-based control.
Identifying the actual system as accurately as possible is the first step of the indirect approach \cite{Lennart1999}, which can be realized by various methods like least squares \cite{AKCAY19931535}, maximum likelihood \cite{UMENBERGER2018280}, set membership \cite{MILANESE1991997}, and subspace methods \cite{McKelvey1996}. As the data is typically noisy, the identified system is uncertain, and one can aim for a controller that robustly stabilizes all the systems within identification error bounds \cite{Mina2020}. However, deriving computationally tractable and tight error bounds from noisy data is a difficult problem in itself \cite{MILANESE1991997, Dean2020}.

The most recent developments on direct data-based control of linear systems can be found in various problems including stabilization \cite{Persis2020}, linear quadratic regulation \cite{Florian2021, Persis2020, Claudio2021A}, $\mathcal{H}_{\infty}$ control \cite{Henk2022, steentjes2021}, model predictive control \cite{Julian2021MPC, Florian2022Predictive}, time delay system control \cite{Juan2021}, networked system control \cite{Baggio2021, Ahmed2021}, linear parameter-varying systems control \cite{Jared2023LPV}, safe control \cite{bisoffi2020controller, luppi2021data} and model reduction \cite{Nima2020}. A theoretic underpinning for most of the aforementioned works is the fundamental lemma by Willems et al. \cite{WILLEMS2005325}. This fundamental lemma shows that, under persistently exciting inputs, a controllable linear system can be equivalently represented by  a finite set of input-state/output samples.

The success of data-driven control methods hinges on the quality/quantity of data. If data is scarcely available or heavily affected by noise, then there is little chance to find an efficient data-driven controller to regulate the behavior of the system. 
The current work is motivated by practical scenarios (examples can be found in \cite{Schoellig2017a, Sundaram2022, Chakrabarty2022, Russo2022}) where collecting data from the \textit{target} system that we wish to identify, optimize, or control is difficult, hazardous, or expensive. The identification or the control task in such cases can rely on the data collected from an accessible similar system/model, which we refer to as the \textit{source} system. 
As a case in point, the target system can be a physical system and the source system can be a numerical model or a digital twin of the target system. Another case in point is where a controller for a (source) system has already been designed, and we want to use the previously collected data to design a controller for a new (target) system knowing that the target system is closely related to the source system, e.g. linearized around a different operating point or both source and target systems belong to a compact set (cf. \cite[Section 5.1]{Carsten2004}, \cite[Section 4.3]{Duan2013}).
The idea that we will pursue here is to leverage the abundant data that can be collected from the source system to control the target system. From a conceptual viewpoint, this idea shares similarities with what is known in the machine learning community as ``transfer learning"  \cite{Fuzhen2021}. Transfer learning essentially amounts to learning a new task through the transfer of knowledge from a related task that has already been learned. 
In the context of dynamical systems, transfer learning is often pursued as finding an optimal transfer map which transforms one dynamical system to another, see e.g. \cite{Botond2013, Schoellig2017a, Schoellig2020a}.
A related problem is to identify one system using data obtained from another system \cite{Sundaram2022, chakrabarty2022a}. 
Such learning methods mainly focus on analysis problems, e.g. finding an optimal transfer map or identifying the target system.
As for the task of controller design, authors in \cite{Chakrabarty2022, Spence2022} collect the source system's data with a pre-designed but possibly poorly tuned controller and then train a better controller for the target system. 
Our transfer-learning method is substantially different from the aforementioned works. Particularly, we do not work with any pre-designed controller, collect open-loop data even though the system is unstable, and assume only very few data samples from the target system. To the best of our knowledge, our note is the first paper that leverages transfer learning for the direct design of stabilizing controllers, providing analytical stability guarantees.

\medskip{}
\subsubsection*{Contribution}

This paper proposes a data-driven method to design a controller that stabilizes the actual target system by using primarily the available data from the source system. Due to scarcity of target system data, stabilizing the set of all systems consistent with the (target) data is infeasible. To cope with this challenge, we use the plenty of data that we have from the source system and find a suitable, and relatively small, compact set of systems that contains the actual target system. To find this set, we assume that the source and target systems are close in a suitable metric. This assumption is in line with the transfer learning idea where the involved systems should be ``related" in some sense. Once we find the aforementioned compact set of systems, we search for a controller that robustly stabilizes all the systems in the set.
We remark that our controller may or may not stabilize the source system, a fact which differentiates the results with what is customary in classical robust control \cite[Section 5.1]{Carsten2004}\cite[Section 4.3]{Duan2013}. For direct data-driven control results concerning a robust stabilization of a set of systems consistent with data, we refer to \cite{Persis2020, Andrea2022Petersen, Henk2020Informativity, Henk2022, Julian2020, berberich2021combining}.

\medskip{}
\subsubsection*{Outline}
Section \ref{sec:Psetup} includes the required notions, definitions, and the problem setup. The main results are provided in Section \ref{sec:results}.
To illustrate better the underlying idea and provide our most explicit results, we first discuss the case where the source system is disturbance-free in Subsection \ref{sec:dis-free}, and accommodate the disturbance in the design later in Subsection \ref{sec:dis-with}. Numerical case studies are provided in Section \ref{sec:Numerical} to demonstrate the theoretical findings. Finally, the note closes with conclusions in Section \ref{sec:Conclusion}.

\section{Problem setup}\label{sec:Psetup}
\subsection{Notation and preliminaries}\label{sec:prelimi}

We denote the set of nonnegative integers by $\mathbb{N}_0$, 
the identity matrix of size $n$ by $I_n$, and the zero matrix of size $m\times n$ by $0_{m \times n}$. The indices are dropped whenever no confusion arises. 
The induced $2-$norm of a matrix $M$ is given by $\| M \| = \sigma_{\max}(M)$, where $\sigma_{\max}(M)$  denotes the largest singular value of $M$.   
For a symmetric matrix $\left[ \begin{smallmatrix} M & N \\ N^{\top} & O \end{smallmatrix} \right]$, we may use the shorthand writing $\left[ \begin{smallmatrix} M & N \\ \ast & O \end{smallmatrix} \right]$. 

A function $f :\, \mathbb{R}^n \to \mathbb{R}$ is called \textit{quadratic} if it admits the form $	f(z) = z^{\top} \mathbf{A} z + z^{\top} \mathbf{B} + \mathbf{B}^{\top} z + \mathbf{C} $ for some matrices $\mathbf{A} = \mathbf{A}^{\top} \in \mathbb{R}^{n \times n}$, $\mathbf{B} \in \mathbb{R}^{n}$ and $\mathbf{C} \in \mathbb{R}$.
A function $f :\, \mathbb{R}^{p \times q} \to \mathbb{R}^{q \times q}$ is called \textit{matrix quadratic} if it admits the form
\begin{equation}\label{eq:quadraticMF}
	f(Z) = Z^{\top} \mathbf{A} Z + Z^{\top} \mathbf{B} + \mathbf{B}^{\top} Z + \mathbf{C}
\end{equation}
for some symmetric matrices $\mathbf{A} \in \mathbb{R}^{p \times p}$, $\mathbf{C} \in \mathbb{R}^{q \times q}$, and a matrix $\mathbf{B} \in \mathbb{R}^{p \times q}$. We frequently use the set of negative semidefinite matrix quadratic functions 
\begin{equation}\label{eq:conic}
	\mathcal{Q} := \left\{ Z \in \mathbb{R}^{p \times q}  :  f(Z) \preceq 0 \right\}. 
\end{equation}
If $\mathbf{A} \succ 0$ and $\mathbf{C} - \mathbf{B}^{\top} \mathbf{A}^{-1} \mathbf{B} \preceq 0$, then the set $\mathcal{Q}$ is compact, providing a natural extension of the classical ellipsoid in the Euclidean space, and is termed here as a hyper-ellipsoid. In case, $\mathbf{C} - \mathbf{B}^{\top} \mathbf{A}^{-1} \mathbf{B} = 0$, the hyper-ellipsoid reduces to a singleton.

\subsection{Transfer Stabilization Problem}\label{sec:problem}
We consider two linear time-invariant (LTI) discrete-time systems
\begin{subequations}\label{eq:system0}
\begin{align}
    x_S(i+1) &= A_{S \star} x_S(i)+B_{S \star} u_S(i)+d_S(i), \ i\in \mathbb{N}_0, \label{eq:sys1} \\
    x_T(i+1) &= A_{T \star} x_T(i)+B_{T \star} u_T(i)+d_T(i), \ i\in \mathbb{N}_0. \label{eq:sys2}
\end{align}	
\end{subequations}
with state $x_S,\ x_T \in \mathbb{R}^n$, input $u_S,\ u_T \in \mathbb{R}^m$, disturbance $d_S,\ d_T \in \mathbb{R}^n$, state matrices $A_{S \star}$, $A_{T \star}$, and input matrices $B_{S \star}$, $B_{T \star}$. The first system \eqref{eq:sys1} is named the source system, and the second system \eqref{eq:sys2} is named the target system. Our goal is to use the source system as a proxy for the target system, and design a controller that can stabilize the actual target system. The distinction between the source and target system originates from the hypothesis that the data/model-based information available on the target system is by far less than that of the source system. 

We pursue a data-driven approach in the sense that we do not treat the actual source system $(A_{S \star}, B_{S \star})$ and the actual target system $(A_{T \star}, B_{T \star})$ to be given a priori, by instead we rely on input-state data obtained through experiments. Consistent with our problem setup, we consider the scenario where the data collected from the source system is much richer than that of the target system.

To collect data, we consider the input sequences 
\[\begin{aligned}
    U_{S0} &:= [u_S(0)\ u_S(1) \cdots u_S(N_S-1)], \\
    U_{T0} &:= [u_T(0)\ u_T(1) \cdots u_T(N_T-1)]
\end{aligned}\]
of length $N_S$ and $N_T$ applied to the source and target systems, respectively. The resulting state response data is collected in the matrices 
\[\begin{aligned}
    X_{S0} &:= [x_S(0)\ x_S(1) \cdots x_S(N_S-1)], \\
    X_{T0} &:= [x_T(0)\ x_T(1) \cdots x_T(N_T-1)].
\end{aligned}\]
The shifted state response matrices are consistently formed as 
\[\begin{aligned}
    X_{S1} &:= [x_S(1)\ x_S(2) \cdots x_S(N_S)], \\
    X_{T1} &:= [x_T(1)\ x_T(2) \cdots x_T(N_T)].
\end{aligned}\]
Analogously, the matrices 
\[\begin{aligned}
    D_{S0} &:= [d_S(0)\ d_S(1) \cdots d_S(N_S-1)], \\
    D_{T0} &:= [d_T(0)\ d_T(1) \cdots d_T(N_T-1)]
\end{aligned}\]
account for the unknown disturbances affecting the source and target data respectively.

Clearly, from \eqref{eq:system0}, we have 
\begin{subequations}\label{eq:system0Mtx}
\begin{align}
    X_{S1} &= A_{S \star} X_{S0}+B_{S \star} U_{S0}+D_{S0}, \\
    X_{T1} &= A_{T \star} X_{T0}+B_{T \star} U_{T0}+D_{T0}. 
\end{align}
\end{subequations}

While the disturbance sequences are considered to be unknown, 
we assume that $D_{S0}$ and $D_{T0}$ have bounded energy, i.e.,
\begin{subequations}\label{eq:D0}
\begin{align}
    D_{S0} & \in \mathcal{D}_S := \left\{ D \in \mathbb{R}^{n \times N_S}  :  D D^{\top} \preceq \Delta_{S} \Delta_{S}^{\top} \right\}, \\
    D_{T0} & \in \mathcal{D}_T := \left\{ D \in \mathbb{R}^{n \times N_T}  :  D D^{\top} \preceq \Delta_{T} \Delta_{T}^{\top} \right\}.
\end{align}
\end{subequations}
where the positive semidefinite matrices $\Delta_{S} \Delta_{S}^{\top}$ and $\Delta_{T} \Delta_{T}^{\top}$ represent our a priori knowledge on the disturbance; see e.g. \cite{Andrea2022Petersen, Julian2020, Henk2022, berberich2021combining} for some recent works where such knowledge has been used. 

Now, bearing in mind \eqref{eq:system0Mtx} and \eqref{eq:D0}, the sets of matrices that are consistent with the source and target data are given by 
\begin{subequations}\label{eq:consistent}
\begin{align}
    \mathcal{E}_S & := \{ (A,B)  :  X_{S1} = A X_{S0} + B U_{S0} + D, \ D  \in  \mathcal{D}_S \}, \label{eq:consistent1}  \\
    \mathcal{E}_T & := \{ (A,B)  :  X_{T1} = A X_{T0} + B U_{T0} + D, \ D  \in  \mathcal{D}_T \}, \label{eq:consistent2}
\end{align}
\end{subequations}
respectively.  

In order to stabilize the actual target system $(A_{T \star}, B_{T \star})$, one may attempt to robustly stabilize all the matrices $(A,B) \in \mathcal{E}_T$, via tools such as S-lemma \cite{Henk2022} or Petersen's lemma \cite{Andrea2022Petersen}. However, the twist here is that the data available on the target system is scarce, thereby making $\mathcal{E}_T$ very large or even unbounded (cf. section \ref{sec:results}). This means that finding a single controller stabilizing the whole set $\mathcal{E}_T$ is very unlikely if not impossible. Our idea instead is to leverage the data of the source system and stabilize a relevant and relatively small subset of $\mathcal{E}_T$ that still guarantees the stabilization of the actual target system $(A_{T \star}, B_{T \star})$.

Contrary to the target system, we assume abundant data can be obtained from the source system. In particular, we assume the following condition, which is satisfied under a persistently exciting input signal \cite{WILLEMS2005325}.

\begin{assumption}\label{asp:rowRank}
The matrix $W_{S0} := \left[\begin{smallmatrix} X_{S0}\\ U_{S0} \end{smallmatrix}\right]$ has full row rank.
\end{assumption}

Obviously, for any sensible transfer of knowledge from the source to the target system, the behavior of the two systems must be {\em close} in some sense. The closeness metric we use here is given by
\begin{equation}\label{eq:close}
    \Big\| [A_{T \star}\ B_{T \star}]-[A_{S \star}\ B_{S \star}] \Big\| \leq \epsilon,   
\end{equation}
where $\epsilon$ is a known positive scalar. This leads to the following assumption:

\begin{assumption}\label{asp:eps-close}
The target system is $\epsilon$-close to the source system, i.e, \eqref{eq:close} holds for some given positive $\epsilon$. 
\end{assumption}

Clearly, an $\epsilon$ satisfying the above condition always exists. Hence, the assumption merely asks for knowing an upper bound of the norm on the left hand side of \eqref{eq:close}. This upper bound is treated as a priori knowledge, and explicitly appears in the proposed design procedure.
Motivated by the preceding discussion, we formulate the following \textit{transfer stabilization} problem:

\begin{problem}\label{plm:problem}
Let Assumption \ref{asp:rowRank} and Assumption \ref{asp:eps-close} hold, and the source and the target system satisfy the data-consistent sets in \eqref{eq:consistent}. Design a state feedback controller $u_T = K x_T$ that makes the closed-loop matrix $A_{T \star} + B_{T \star} K$ Schur stable.
\end{problem}

\section{Main results}\label{sec:results}

This section reports the main results of the note. We first discuss the case where the data of the source system is clean, i.e. the effect of disturbance on the data is neglected. Then, we provide the extension to the more general case where the source data is affected by the disturbance. Note that in both cases the disturbance is present on the limited data obtained from the target system. 

To prepare for the subsequent results, we  rewrite the data-consistent sets \eqref{eq:consistent} in the following matrix quadratic form:
\begin{subequations}\label{eq:systemset}
\begin{align}
    \mathcal{E}_S =  \{ & (A,B)  :  Z = [A\ B]^{\top}, \nonumber \\  
    & Z^{\top} \mathbf{A}_S Z  +  Z^{\top} \mathbf{B}_S  +  \mathbf{B}_S^{\top} Z  +  \mathbf{C}_S
    :=f_S(Z) \preceq 0 \}, \label{eq:systemset1} \\
    \mathcal{E}_T =  \{ & (A,B)  :  Z = [A\ B]^{\top}, \nonumber \\  
    & Z^{\top} \mathbf{A}_T Z  +  Z^{\top} \mathbf{B}_T  +  \mathbf{B}_T^{\top} Z  +  \mathbf{C}_T
    :=f_T(Z)  \preceq 0 \}, \label{eq:systemset2}
\end{align}
\end{subequations}
where
\begin{subequations}\label{eq:systemsetABC}
\begin{align}
    \mathbf{A}_S = & W_{S0} W_{S0}^{\top}, \quad  \mathbf{B}_S = -W_{S0} X_{S1}^{\top}, \nonumber \\
    & \mathbf{C}_S = -\Delta_S \Delta_S^{\top}+X_{S1} X_{S1}^{\top}; \label{eq:systemsetABC-1}  \\
    \mathbf{A}_T = & W_{T0} W_{T0}^{\top}, \quad  \mathbf{B}_T = -W_{T0} X_{T1}^{\top}, \nonumber \\
    & \mathbf{C}_T = -\Delta_T \Delta_T^{\top}+X_{T1} X_{T1}^{\top}, \label{eq:systemsetABC-2}
\end{align}
\end{subequations}
and $ W_{T0} := \left[ \begin{array}{c} X_{T0} \\ U_{T0} \end{array} \right] $.

Under Assumption \ref{asp:rowRank}, $\mathcal{E}_S$ becomes a compact set \cite[Lemma 2]{Andrea2022Petersen} since $ \mathbf{A}_S \succ 0 $ and $ \mathbf{C}_S - \mathbf{B}_S^{\top} \mathbf{A}_S^{-1} \mathbf{B}_S \preceq 0 $.
The former inequality directly follows from the assumption whereas the latter follows from a Schur complement argument \cite[Lemma 1]{Andrea2022Petersen}.
On the contrary, $\mathcal{E}_T$ is not in general compact due to the limited data available on the target system, which leads to a singular $\mathbf{A}_T$. Note that boundedness of $\mathcal{E}_T$ and nonsingularity of $\mathbf{A}_T$ requires at least $n+m$ data samples.

\subsection{Disturbance-free source system}\label{sec:dis-free} 

\begin{figure}[ht]
\begin{center}
    \includegraphics[scale=0.9]{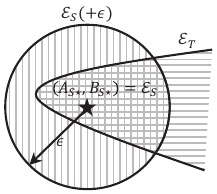}
    \caption{Geometric illustration for scalar source and target systems.
    The center of the disk (i.e., the solid star) is the data consistent set $\mathcal{E}_S$ which in disturbance-free case is a singleton given by the actual source system $(A_{S \star}, B_{S \star})$. The disk is $\mathcal{E}_S(+\epsilon)$ which is the set of all possible target systems that are $\epsilon$-close to the actual source system. Because of the limited data available from the target system, $\mathcal{E}_T$ may be very large or even unbounded.}\label{fig:NoiseFreeCase0}
\end{center}	
\end{figure}

In the case the data from the source system is disturbance free, the set $\mathcal{E}_S$ in \eqref{eq:consistent1} (equivalently, \eqref{eq:systemset1}) reduces to the singleton $(A_{S \star}, B_{S \star})$ which is the actual source system. This is a direct consequence of Assumption \ref{asp:rowRank} and the definition of the set $\mathcal{E}_S$.
As for the actual target system, we know that it belongs to the data-consistent set $\mathcal{E}_T$ and is $\epsilon$-close to the source system, namely \eqref{eq:close} holds.
In other words, we have
\begin{equation}\label{eq:capp}
    (A_{T \star}, B_{T \star}) \in \mathcal{E}_T \cap \mathcal{E}_S(+\epsilon),
\end{equation}
where $\mathcal{E}_S(+\epsilon)$ is the set of systems that are $\epsilon$-close to the source system, i.e,
\begin{equation}\label{eq:CoveredSet}
    \mathcal{E}_S(+\epsilon) := \left\{ (A,B) : \Big\| [A\ B] - [A_{S \star}\ B_{S \star}] \Big\|  \leq \epsilon \right\}.
\end{equation}
As the actual target system is unknown, we aim at stabilizing the set of systems in the intersection given by the right hand side of \eqref{eq:capp}. This is visualized in the special case of scalar systems in Figure \ref{fig:NoiseFreeCase0}. 

\begin{figure*}[!t]


\setcounter{mytempeqncnt}{\value{equation}}
\setcounter{equation}{16}

\begin{equation}\label{eq:sufficientController}
    \left[\begin{array}{cccc} P-\beta I & 0 & 0 & 0 \\ \ast & -P & -Y^{\top} & 0 \\ \ast  & \ast  & 0 & Y \\ \ast  & \ast  & \ast & P \end{array}\right]       -\tau_T \left[\begin{array}{ccc} -\mathbf{C}_T & -\mathbf{B}_T^{\top} & 0 \\ \ast & -\mathbf{A}_T & 0 \\ \ast & \ast & 0 \end{array}\right]       -\tau_S \left[\begin{array}{ccc} -\mathbf{C}_{\epsilon} & -\mathbf{B}_{\epsilon}^{\top} & 0 \\ \ast & -\mathbf{A}_{\epsilon} & 0 \\ \ast & \ast & 0 \end{array}\right]  \succeq 0, 
\end{equation}
\setcounter{equation}{\value{mytempeqncnt}}

\vspace*{4pt}
\hrulefill

\end{figure*}

\noindent \textbf{\textit{Stabilizing controller.}} Now, we aim to find a controller gain matrix $K$ that stabilizes all systems belonging to the intersection in \eqref{eq:capp}. 
This is formally formulated as
\begin{subequations}\label{eq:problem2solve}
\begin{align}
    \text{find} \quad & P \succ 0, \quad \beta > 0, \quad K  \label{eq:problem2solve-1} \\
    \text{s.t.} \quad & (A+BK)P(A+BK)^{\top}-P \preceq -\beta I, \label{eq:problem2solve-2} \\ 
    & \forall(A, B)\in \mathcal{E}_T \cap \mathcal{E}_S(+\epsilon), \label{eq:problem2solve-3}
\end{align}
\end{subequations}
where \eqref{eq:problem2solve-2} is the Lyapunov stability inequality in discrete time. 

Recall that $\mathcal{E}_T$ is identified by the quadratic matrix inequality $f_T(Z)\preceq 0$ given by \eqref{eq:systemset2}. 
Moreover, we can rewrite \eqref{eq:CoveredSet} in a matrix quadratic form as:
\begin{equation}\label{eq:NoisFreeCase}
\begin{aligned}
    & \mathcal{E}_S(+\epsilon) = \{ (A,B) : Z = [A\ B]^{\top}, \\  
    & \quad \quad \quad \quad \ Z^{\top} \mathbf{A}_{\epsilon} Z + Z^{\top} \mathbf{B}_{\epsilon} + \mathbf{B}_{\epsilon}^{\top} Z + \mathbf{C}_{\epsilon} := f_{\epsilon}(Z) \preceq 0 \},
\end{aligned}
\end{equation}
where 
\begin{align}
    \mathbf{A}_{\epsilon} = I, \quad & \mathbf{B}_{\epsilon} = -Z_{cS}, \label{eq:NoisFreeCase-a} \tag{13a}  \\
    \mathbf{C}_{\epsilon} = Z_{cS}^{\top} Z_{cS} - \epsilon^2 I, \quad &  Z_{cS} = [A_{S \star}\ B_{S \star}]^{\top}. \label{eq:NoisFreeCase-b} \tag{13b} 
\end{align}
Finally, the Lyapunov stability inequality \eqref{eq:problem2solve-2} can also be stated as a quadratic matrix inequality (see also \cite{Andrea2022Petersen, Henk2022}):
\begin{equation}\label{eq:controlquad}
    Z^{\top} \mathbf{A}_c Z + Z^{\top} \mathbf{B}_c + \mathbf{B}_c^{\top} Z + \mathbf{C}_c := f_c(Z) \preceq 0,
\end{equation}
where 
\begin{equation}\label{eq:AcBcCc}
    \mathbf{A}_c  =  \left[ \begin{matrix} P & PK^{\top} \\ \ast & KPK^{\top} \end{matrix} \right], \ \mathbf{B}_c  =  0_{(n+m) \times n}, \ \mathbf{C}_c  =  -P + \beta I,
\end{equation}
and $Z = [A\ B]^{\top}$ is the independent input argument of  $f_c(Z)$.

By using the aforementioned quadratic matrix inequalities, the stabilization problem \eqref{eq:problem2solve} can be recast as 
\begin{subequations}\label{eq:problem2solveV}
\begin{align}
    \text{find} \quad & P \succ 0, \quad \beta > 0, \quad K, \label{eq:problem2solveV-1} \\ 
    \text{s.t.} \quad & f_c(Z) \preceq 0, \ \forall Z \in \mathcal{I}:=\{Z :  f_T(Z) \preceq 0 \nonumber \\ 
	& \quad \quad \quad \quad \quad \quad \quad \quad \quad \quad \quad \quad  {\text{ and }} f_{\epsilon}(Z) \preceq 0\}, \label{eq:problem2solveV-2}
\end{align}
\end{subequations}
where $f_T(Z)$, $f_\epsilon(Z)$, and $f_c(Z)$ are given by \eqref{eq:systemset2}, \eqref{eq:NoisFreeCase}, and \eqref{eq:controlquad}, respectively. The following result provides an LMI condition solving the stabilization problem \eqref{eq:problem2solveV}:

\medskip{}
\begin{theorem}\label{thm:sufficientController} 
Suppose that \eqref{eq:sufficientController}\footnote{See the matrix inequality at the top of the next page, where matrices $\mathbf{A}_T$, $\mathbf{B}_T$, $\mathbf{C}_T$ are defined in \eqref{eq:systemsetABC-2} and $\mathbf{A}_{\epsilon}$, $\mathbf{B}_{\epsilon}$, $\mathbf{C}_{\epsilon}$ are defined in \eqref{eq:NoisFreeCase-a}, \eqref{eq:NoisFreeCase-b}.} holds for some matrix $P \succ 0$ and scalars $\beta > 0, \tau_S \geq 0, \tau_T \geq 0$.
Then, the controller $K = YP^{-1}$ solves the stabilization problem \eqref{eq:problem2solveV} (equivalently, \eqref{eq:problem2solve}). In particular, $K$ stabilizes the target system, i.e, $A_{T \star}+B_{T \star} K$ is Schur stable. 
\end{theorem}

\setcounter{mytempeqncnt}{17}
\setcounter{equation}{\value{mytempeqncnt}}

\begin{pfof} {Theorem \ref{thm:sufficientController}}
Suppose that \eqref{eq:sufficientController} is satisfied. Let $Y=KP$.
By using a Schur complement argument, we can equivalently write \eqref{eq:sufficientController} as
\begin{equation}\label{eq:suffiCtrlInter}
    \left[\begin{array}{cc} \mathbf{C}_c & \mathbf{B}_c^{\top} \\ \ast & \mathbf{A}_c \end{array}\right]       -\tau_T \left[\begin{array}{cc} \mathbf{C}_T & \mathbf{B}_T^{\top} \\ \ast & \mathbf{A}_T \end{array}\right]        -\tau_S \left[\begin{array}{cc} \mathbf{C}_{\epsilon} & \mathbf{B}_{\epsilon}^{\top} \\ \ast & \mathbf{A}_{\epsilon} \end{array}\right]    \preceq 0,
\end{equation}
where $\mathbf{A}_c$, $\mathbf{B}_c$, and $\mathbf{C}_c$ are given by \eqref{eq:AcBcCc}. 
By the lossy matrix S-procedure in \cite[Lemma 2]{Andrea2021TradeOffs}, the above inequality implies that \eqref{eq:problem2solveV-2} holds, namely the matrix quadratic inequality $f_c(Z) \preceq 0$ holds for all $Z\in \mathcal{I}$. 
Hence, $K=YP^{-1}$ solves the stabilization problem \eqref{eq:problem2solveV} and, equivalently, \eqref{eq:problem2solve}. Noting \eqref{eq:capp}, the latter implies that $(A_{T \star}, B_{T \star})$ is Schur stable.
\end{pfof}

In general, the LMI in \eqref{eq:sufficientController} is only sufficient to solve the stabilization problem \eqref{eq:problem2solveV} (equivalently, \eqref{eq:problem2solve}).
The next result shows that under additional technical conditions, the LMI in \eqref{eq:sufficientController} becomes necessary and sufficient. 

\begin{proposition}\label{pp:NeceSuff}
Assume that
\begin{equation}\label{eq:NeceSuffiSuppsose}
    \mu_T \left[\begin{array}{cc} \mathbf{C}_T & \mathbf{B}_T^{\top} \\ \ast & \mathbf{A}_T \end{array}\right]        +\mu_S \left[\begin{array}{cc} \mathbf{C}_{\epsilon} & \mathbf{B}_{\epsilon}^{\top} \\ \ast & \mathbf{A}_{\epsilon} \end{array}\right]   \succ  0,
\end{equation}
for some $\mu_T, \mu_S \in \mathbb{R}$. 
{Moreover, assume that there exists a matrix 
$\bar{Z}\in \mathcal{E}_T \cap \mathcal{E}_S(+\epsilon)$ such that  $f_T(\bar{Z})$ and $ f_{\epsilon}(\bar{Z})$ are both nonsingular.\footnote{See Remark \ref{rmk:noneat}.}
}Then, there exist $P \succ 0$, $\beta > 0, \tau_S \geq 0, \tau_T \geq 0$ satisfying \eqref{eq:sufficientController} if and only if there exist $P \succ 0$, $\beta > 0$, and $K$ satisfying \eqref{eq:problem2solveV-2}.
\end{proposition}

To prove the result of the above proposition, we need the following lemma, whose proof requires a few additional results and is provided in Appendix. 
\begin{lemma}\label{lm:Matrix-S-S} 
Let $f_i : \mathbb{R}^{p \times q} \to \mathbb{R}^{q \times q}$, $p> 1$, $i = 0,1,2$, be matrix quadratic functions admitting the form  in \eqref{eq:quadraticMF}, i.e., $f_i(Z) = Z^{\top} \mathbf{A}_i Z + Z^{\top} \mathbf{B}_i + \mathbf{B}_i^{\top} Z + \mathbf{C}_i$. Let  
$\mathbf{A}_{Qi} := \left[ \begin{smallmatrix} \mathbf{C}_i & \mathbf{B}_i^{\top} \\ \mathbf{B}_i & \mathbf{A}_i \end{smallmatrix} \right]$ for each $i$. Suppose that there exist $\mu_1, \mu_2 \in \mathbb{R}$, $\bar{Z}\in\mathbb{R}^{p \times q}$ such that
\begin{align}
     \mu_1 \mathbf{A}_{Q1} + \mu_2 \mathbf{A}_{Q2} \succ 0, \label{eq:AsmpTheorem1-1} \\
     f_1(\bar{Z}) \prec 0,\ f_2(\bar{Z}) \prec 0. \label{eq:AsmpTheorem1-2}
\end{align}
Then, the following statements are equivalent:
\begin{itemize}
    \item[(I)] $f_0(Z) \preceq 0$, $\forall Z : f_i(Z) \preceq 0, \, i=1,2$.
    \item[(II)] $f_0(Z) \preceq 0$, $\forall Z : f_i(Z) \prec 0, \, i=1,2$.
    \item[(III)] $\exists$ $\tau_1 \geq 0$, $\tau_2 \geq 0$ such that $\mathbf{A}_{Q0} \preceq \tau_1 \mathbf{A}_{Q1} + \tau_2 \mathbf{A}_{Q2}$.
\end{itemize}
\end{lemma}


\begin{pfof} {Proposition \ref{pp:NeceSuff}}
First, we resort to the result of Lemma \ref{lm:Matrix-S-S} and choose $f_0(Z) = f_c(Z)$, $f_1(Z) = f_T(Z)$, $f_2(Z) = f_{\epsilon}(Z)$. Note that \eqref{eq:AsmpTheorem1-1} holds due to \eqref{eq:NeceSuffiSuppsose}.  {Moreover, as $\bar{Z} \in \mathcal{E}_T \cap \mathcal{E}_S(+\epsilon)$, we have $f_T(\bar{Z}) \preceq 0$ and $f_{\epsilon}(\bar{Z}) \preceq 0$. The latter together with the nonsingularity assumption of $f_T(\bar{Z})$ and $f_{\epsilon}(\bar{Z})$ result in $f_T(\bar{Z}) \prec 0$ and $f_{\epsilon}(\bar{Z}) \prec 0$, which verifies the condition \eqref{eq:AsmpTheorem1-2}.}
Now by using the equivalence between the first and the third statement of Lemma \ref{lm:Matrix-S-S}, we  
conclude that the matrix inequality in \eqref{eq:problem2solveV-2} is equivalent to the one in \eqref{eq:suffiCtrlInter}. 
The equivalence between the matrix inequalities \eqref{eq:suffiCtrlInter} and \eqref{eq:sufficientController} follows from the Schur complement argument and the relation $Y=KP$. This completes the proof. 
\end{pfof}

\begin{remark} 
Lemma \ref{lm:Matrix-S-S} provides an extension of \cite[Theorem 4.1]{Polyak1998} from quadratic vector functions to quadratic matrix functions. We refer to \cite[Lemma 2]{Andrea2021TradeOffs} (lossy matrix S-procedure) for the implication (III) $\Rightarrow$ (I), and to \cite{Henk2022} (matrix S-lemma) for the case where $f_1(Z)=f_2(Z)$.  
\end{remark}
	
\begin{remark}\label{rmk:noneat} 
Note that \eqref{eq:NeceSuffiSuppsose} in Proposition \ref{pp:NeceSuff} descends from the condition in \eqref{eq:AsmpTheorem1-1}. 
Moreover, the set $\mathcal{E}_T \cap \mathcal{E}_S(+\epsilon)$ is nonempty due to \eqref{eq:capp}. The technical condition on nonsingularity of $f_T(\bar{Z})$ and $f_{\epsilon}(\bar{Z})$ enables us to verify \eqref{eq:AsmpTheorem1-2}, which serves as a Slater-type condition; see also \cite[Theorem 9]{Henk2022} and \cite[Fact 3]{Andrea2022Petersen}.	
\end{remark}

\subsection{Disturbance on both source and target systems}\label{sec:dis-with} 

Recall that the actual target system belongs to the data-consistent set $\mathcal{E}_T$ and is also $\epsilon$-close to the source system, namely \eqref{eq:close} holds.
However, in case the source system data is affected by disturbances, the data-consistent set $\mathcal{E}_S$ in \eqref{eq:systemset1} is no longer a singleton but a matrix ellipsoid. Consequently, we can no longer find a data-based characterization of the intersection on the right hand side of \eqref{eq:capp} and solve the corresponding stabilization problem \eqref{eq:problem2solve}. 

The first observation that we make here is that the actual target system is at most $\epsilon$-away from the source system data-consistent set $\mathcal{E}_S$, namely 
\begin{equation}\label{eq:CoveredSetDis}
\begin{aligned}
    & (A_{T \star},  B_{T \star}) \in \Upsilon_S(\epsilon) := \Big\{ (A,B) : \Big\| [A\ B]  \\
    & \quad \quad \quad \quad \quad \ - [A_S\ B_S] \Big\|  \leq \epsilon, \ \text{for some}  \left( A_S, B_S \right) \in \mathcal{E}_S  \Big\}.
\end{aligned}
\end{equation}
The above holds due to \eqref{eq:close} and the fact that $ (A_{S \star}, B_{S \star}) \in \mathcal{E}_S$.

\begin{figure}[H]
\begin{center}
    \includegraphics[scale=0.9]{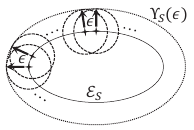}
    \caption{Geometric illustration for $\Upsilon_S(\epsilon)$ in scalar systems.}\label{fig:Union}
\end{center}	
\end{figure}

The set $\Upsilon_S(\epsilon)$ is visualized in Figure \ref{fig:Union} for the special case of scalar systems. In this case, $\Upsilon_S(\epsilon)$ is the union of $\mathcal{E}_S$ and all
disks with radius $\epsilon$ centered on the boundary points of the ellipsoid $\mathcal{E}_S$.

\begin{figure*}[!t]

\setcounter{mytempeqncnt}{\value{equation}}
\setcounter{equation}{27}

\begin{equation}\label{eq:sufficientControllerNoise}
    \left[\begin{array}{cccc} P-\beta I & 0 & 0 & 0 \\ \ast & -P & -Y^{\top} & 0 \\ \ast  & \ast  & 0 & Y \\ \ast  & \ast  & \ast & P \end{array}\right]         -\tau_T \left[\begin{array}{ccc} -\mathbf{C}_T & -\mathbf{B}_T^{\top} & 0 \\ \ast & -\mathbf{A}_T & 0 \\ \ast & \ast & 0 \end{array}\right]       -\tau_S \left[\begin{array}{ccc} -\mathbf{C}_{{\Upsilon}} & -\mathbf{B}_{{\Upsilon}}^{\top} & 0 \\ \ast & -\mathbf{A}_{{\Upsilon}} & 0 \\ \ast & \ast & 0 \end{array}\right]  \succeq 0, 
\end{equation}
\setcounter{equation}{\value{mytempeqncnt}}

\vspace*{4pt}
\hrulefill
    
\end{figure*}

We then attempt to follow the footsteps of the data-based stabilization algorithm discussed in Subsection \ref{sec:dis-free} with $\mathcal{E}_S(+\epsilon)$ being replaced by $\Upsilon_S(\epsilon)$. 
However, this is not feasible since $\Upsilon_S(\epsilon)$ cannot in general be expressed as a quadratic matrix inequality. To overcome this challenge, we work with an outer-approximation of $\Upsilon_S(\epsilon)$ that can be represented by a quadratic matrix inequality. 

To this end, we first introduce another equivalent expression of $\mathcal{E}_S$:
\begin{equation}\label{eq:r_systemset}
    \begin{aligned}
    \mathcal{E}_S =& \{(A,B) : Z = [A\ B]^{\top}, \\ 
    & \quad \quad \quad \quad \left( Z - Z_{cS} \right)^{\top} \mathbf{A}_S  \left( Z - Z_{cS} \right) \preceq \mathbf{Q}_S \},
\end{aligned}
\end{equation}
where
\[ Z_{cS} = -\mathbf{A}_S^{-1}\mathbf{B}_S,  \quad \mathbf{Q}_S = \mathbf{B}_S^{\top} \mathbf{A}_S^{-1} \mathbf{B}_S - \mathbf{C}_S,\]
with $\mathbf{A}_S$, $\mathbf{B}_S$ and $\mathbf{C}_S$ are given by \eqref{eq:systemsetABC-1}.

\begin{figure}[H]
\begin{center}
    \includegraphics[scale=0.9]{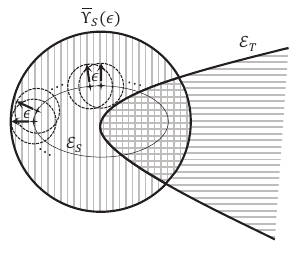}
    \caption{Geometric illustration for scalar source and target systems; both with disturbance.}\label{fig:UnionWithNoise} 
\end{center}	
\end{figure}

The next lemma provides a convex outer approximation of $\Upsilon_S(\epsilon)$ that admits a quadratic matrix inequality representation; see Figure \ref{fig:UnionWithNoise} for the case of scalar systems. The result considers spherical outer-approximations; any other convex outer-approximation of  $\Upsilon_S(\epsilon)$ can be used as long as it can be written as a quadratic matrix inequality.

\begin{lemma}\label{lm:overEpplisoid_backup}
Consider the spectral decomposition $\mathbf{A}_S=U_S^{\top} \Lambda_S U_S$ where $U_S$ is a unitary matrix and $\Lambda_S$ is a positive definite diagonal matrix. Let 
\[r := \| \Lambda_S^{-1/2} \| \| \mathbf{Q}_S^{1/2} \| + \epsilon.\]
Define
\begin{equation}\label{eq:overExpression}
    \begin{aligned}
        \overline{\Upsilon}_S(\epsilon) & := \{ (A,B) : Z = [A\ B]^{\top}, \\  
        & Z^{\top} \mathbf{A}_{{\Upsilon}} Z + Z^{\top} \mathbf{B}_{{\Upsilon}} + \mathbf{B}_{{\Upsilon}}^{\top} Z + \mathbf{C}_{{\Upsilon}} :=f_{{\Upsilon}}(Z) \preceq 0 \},
    \end{aligned}
\end{equation}
with 
\begin{align}
    \mathbf{A}_{{\Upsilon}}  =  I, \quad   \mathbf{B}_{{\Upsilon}} =  - Z_{cS}, \quad \mathbf{C}_{{\Upsilon}}  =  Z_{cS}^{\top}  Z_{cS} - r^2I. \label{eq:overExpression-a} \tag{24a}
\end{align}
Then, ${\Upsilon}_S(\epsilon) \subseteq \overline{\Upsilon}_S(\epsilon)$, where ${\Upsilon}_S(\epsilon)$ is given by \eqref{eq:CoveredSetDis}. 
\end{lemma}

\begin{pfof} {Lemma \ref{lm:overEpplisoid_backup}}
By \eqref{eq:CoveredSetDis}, for any ${Z}_{\Upsilon} \in {\Upsilon}_S(\epsilon)$, there exists ${Z}_{\Upsilon}^{\prime} \in \mathcal{E}_S$ such that $\| {Z}_{\Upsilon} - {Z}_{\Upsilon}^{\prime} \| \leq \epsilon$. Moreover, it is easy to verify that $\| {Z}_{\Upsilon}^{\prime} - Z_{cS} \| \leq \| \Lambda_S^{-1/2} \| \| \mathbf{Q}_S^{1/2} \|$; see the proof of \cite[Lemma 2]{Andrea2022Petersen}. By using the triangle inequality, we have
\[ \| {Z}_{\Upsilon} - Z_{cS} \| \leq \| {Z}_{\Upsilon} - {Z}_{\Upsilon}^{\prime} \| + \| {Z}_{\Upsilon}^{\prime} - Z_{cS} \| \leq \epsilon + \| \Lambda_S^{-1/2} \| \| \mathbf{Q}_S^{1/2} \| = r. \]
This implies that for any ${Z}_{\Upsilon} \in {\Upsilon}_S(\epsilon)$, it holds that
\[ ({Z}_{\Upsilon} - Z_{cS})^{\top} ({Z}_{\Upsilon} - Z_{cS}) \preceq r^2I. \]
By expanding the above expression, we obtain that $f_{{\Upsilon}}({Z}_{\Upsilon}) \preceq 0$ and thus ${Z}_{\Upsilon} \in \overline{\Upsilon}_S(\epsilon)$.
This completes the proof.
\end{pfof}

Now by leveraging the matrix quadratic set $\overline{\Upsilon}_S(\epsilon)$ in \eqref{eq:overExpression}, we can follow a similar procedure as in Section \ref{sec:dis-free}. In this case, we replace \eqref{eq:capp} by 
\begin{equation}\label{eq:cappRsOver}
    (A_{T \star}, B_{T \star}) \in \mathcal{E}_T \cap \overline{\Upsilon}_S(\epsilon).
\end{equation}
As the actual target system is unknown, we aim at stabilizing the set of systems in the intersection given by the right hand side of \eqref{eq:cappRsOver}. This is visualized in the special case of scalar systems in Figure \ref{fig:UnionWithNoise}.

\medskip{}
\noindent \textbf{\textit{Stabilizing controller.}} We aim to find a controller gain matrix $K$ that stabilizes all systems belonging to the intersection in \eqref{eq:cappRsOver}. 
This is formally formulated as
\begin{subequations}\label{eq:problem2solveNoise}
\begin{align}
    \text{find} \quad & P \succ 0, \quad \beta > 0, \quad K  \label{eq:problem2solveNoise-1} \\
    \text{s.t.} \quad & (A+BK)P(A+BK)^{\top}-P \preceq -\beta I, \label{eq:problem2solveNoise-2} \\ 
    & \forall(A, B)\in \mathcal{E}_T \cap \overline{\Upsilon}_S(\epsilon). \label{eq:problem2solveNoise-3}
\end{align}
\end{subequations}

By using the aforementioned quadratic matrix inequalities, the stabilization problem \eqref{eq:problem2solveNoise} can be recast as 
\begin{subequations}\label{eq:problem2solveNoiseV}
\begin{align}
    \text{find} \quad & P \succ 0, \quad \beta > 0, \quad K, \label{eq:problem2solveNoiseV-1} \\ 
    \text{s.t.} \quad & f_c(Z) \preceq 0,\ \forall Z \in \mathcal{I}:=\{Z :  f_T(Z) \preceq 0 \nonumber \\
	& \quad \quad \quad \quad \quad \quad \quad \quad \quad \quad \quad \quad {\text{ and }} f_{{\Upsilon}}(Z) \preceq 0\}, \label{eq:problem2solveNoiseV-2}
\end{align}
\end{subequations}
where $f_T(Z)$, $f_{{\Upsilon}}(Z)$, and $f_c(Z)$ are given by \eqref{eq:systemset2}, \eqref{eq:overExpression}, and \eqref{eq:controlquad}, respectively.

The following result provides an LMI condition solving the stabilization problem \eqref{eq:problem2solveNoiseV}:
\begin{theorem}\label{thm:sufficientControllerNoise} 
Suppose that \eqref{eq:sufficientControllerNoise}\footnote{See the matrix inequality at the top of this page, where matrices $\mathbf{A}_T$, $\mathbf{B}_T$, $\mathbf{C}_T$ are defined in \eqref{eq:systemsetABC-2} and $\mathbf{A}_{{\Upsilon}}$, $\mathbf{B}_{{\Upsilon}}$, $\mathbf{C}_{{\Upsilon}}$ are defined in \eqref{eq:overExpression-a}.} holds for some matrix $P \succ 0$ and scalars $\beta > 0, \tau_S \geq 0, \tau_T \geq 0$.
Then, the controller $K = YP^{-1}$ solves the stabilization problem \eqref{eq:problem2solveNoiseV} (equivalently, \eqref{eq:problem2solveNoise}). In particular, $K$ stabilizes the target system, i.e, $A_{T \star}+B_{T \star} K$ is Schur stable. 
\end{theorem}

\setcounter{mytempeqncnt}{28}
\setcounter{equation}{\value{mytempeqncnt}}

\begin{pfof} {Theorem \ref{thm:sufficientControllerNoise}}
The proof is analogous to the proof of Theorem \ref{thm:sufficientController}, while the only difference is changing $\mathbf{A}_{\epsilon}$, $\mathbf{B}_{\epsilon}$ and $\mathbf{C}_{\epsilon}$ in \eqref{eq:sufficientController} into $\mathbf{A}_{{\Upsilon}}$, $\mathbf{B}_{{\Upsilon}}$ and $\mathbf{C}_{{\Upsilon}}$ in \eqref{eq:sufficientControllerNoise}, respectively.
\end{pfof}

\begin{remark} 
The LMI in \eqref{eq:sufficientControllerNoise} is sufficient to solve the stabilization problem \eqref{eq:problem2solveNoiseV} (equivalently, \eqref{eq:problem2solveNoise}).
For a necessary and sufficient condition one can restate the result of Proposition \ref{pp:NeceSuff} with $\mathbf{A}_{\epsilon}$, $\mathbf{B}_{\epsilon}$, $\mathbf{C}_{\epsilon}$, and $\mathcal{E}_S(+\epsilon)$ being replaced by $\mathbf{A}_{{\Upsilon}}$, $\mathbf{B}_{{\Upsilon}}$, $\mathbf{C}_{{\Upsilon}}$, and $ \overline{\Upsilon}_S(\epsilon)$, respectively. {Note that the LMI \eqref{eq:sufficientControllerNoise}, due to the outer-approximation, is in any case only sufficient for stabilizing the set of systems in $\mathcal{E}_T \cap {\Upsilon}_S(\epsilon).$ }
\end{remark}

\begin{remark}
Note that Theorem \ref{thm:sufficientController} and Theorem \ref{thm:sufficientControllerNoise} solve Problem \ref{plm:problem} formulated in Section \ref{sec:problem}, namely they guarantee that the closed-loop of the target system, i.e. $A_{T \star}+B_{T \star}K$, is Schur stable. Although we accounted for the disturbances only in the open-loop data collection phase, and not explicitly during the controller execution phase, Schur stability of $A_{T \star} + B_{T \star} K$ readily implies input-to-state stability \cite[p.~171]{SontagISS2008}, and thus the state of the controlled target system remains in a neighborhood of the origin in case bounded disturbances are present during the execution of the controller.  
\end{remark}

\section{Case Studies}\label{sec:Numerical} 
\subsection{First order system}
Suppose the source system and target system admit the first-order dynamics 
\[x(i+1) = A x(i) + B u(i) + d(i), \quad i\in \mathbb{N}_0.\]
For source system, we first consider the scenario where it is disturbance free, hence its dynamics are known, which we set as $A_{S \star} = 1.021$ and $B_{S \star} = 0.041$, leading to an unstable system. As for the target system, we set $\epsilon = 0.025$, i.e., the actual target system is $0.025$-close to the source system. We select one instance $(A_{T \star}, B_{T \star})$ arbitrarily from $\mathcal{E}_S(+\epsilon)$ as the true (unknown) target system and apply the input $u \thicksim U[-10,10]$ and the disturbance $d \thicksim U[-1,1]$, where $U[a, b]$ denotes random variables uniformly distributed in $[a, b]$. Note that the value of $(A_{T \star}, B_{T \star})$ is only used to generate data, but not to design the controller. The data length is $N_T = 1$, a single snapshot from input-state data of the target system is available. Bearing in mind \eqref{eq:D0}, the upper bound on the disturbance (i.e., 1) and the data length (i.e., 1), we have $\Delta_{T} \Delta_{T}^{\top} = 1$.

\begin{figure}[t]
\begin{center}
    \includegraphics[width=0.55\columnwidth]{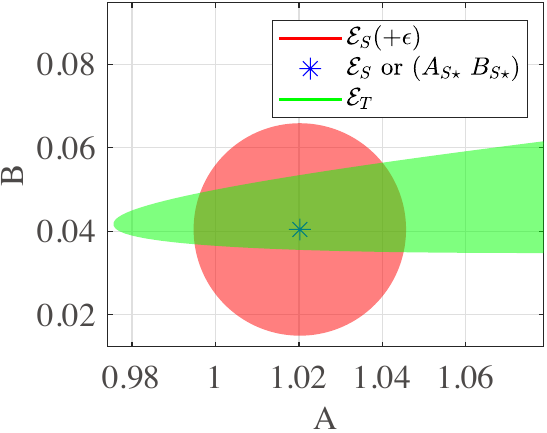}
    \vspace{-0.2cm}
    \caption{First order systems; source system without disturbance; the set $\mathcal{E}_{S}$ coincides with the singleton $(A_{S \star}, B_{S \star})$; the red area depicts the set $\mathcal{E}_{S}(+\epsilon)$ in \eqref{eq:CoveredSet}; the green area shows the set $\mathcal{E}_{T}$ in \eqref{eq:consistent2}. }\label{fig:FirstOrder_DisFree}
\end{center}	
\end{figure}

The key sets in the proposed transfer stabilization scheme are depicted in Figure \ref{fig:FirstOrder_DisFree}. As we can see, stabilizing the intersection $\mathcal{E}_{T} \cap \mathcal{E}_S(+\epsilon)$ is much less conservative than stabilizing all the systems belonging to the set $\mathcal{E}_{T}$. Moreover, solving Theorem \ref{thm:sufficientController} by CVX \cite{cvx}, we have $K = -17.76$ that stabilizes all the system matrices $(A,B)$ contained in the intersection, which implies that the closed-loop matrix $A_{T \star} + B_{T \star} K$ is Schur stable.

Next, we consider the case where the source system is affected by disturbance. We apply the input $u \thicksim U[-10,10]$ and disturbance $d \thicksim U[-0.1,0.1]$ to obtain the data from the source system. The data length is $N_S = 200$. Bearing in mind \eqref{eq:D0}, the upper bound on the disturbance (i.e., 0.1) and the data length (i.e., 200), we have $\Delta_{S} \Delta_{S}^{\top} = 2$. For the target system, we use the same data point as before. In this case, we choose $\overline{\Upsilon}_S(\epsilon)$ as discussed in Lemma \ref{lm:overEpplisoid_backup}.

\begin{figure}[t]
\begin{center}
    \includegraphics[width=0.55\columnwidth]{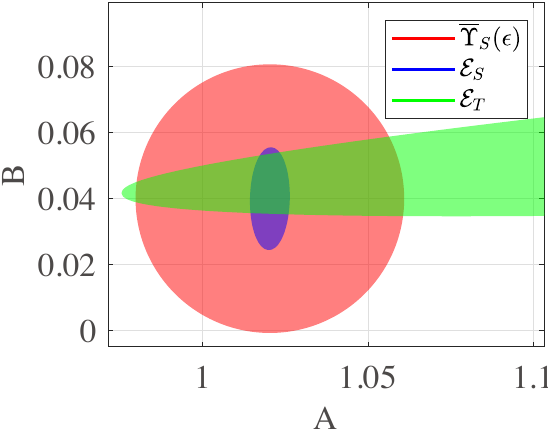}
    \vspace{-0.2cm}
    \caption{First order systems; source system with disturbance; the blue area depicts the set $\mathcal{E}_{S}$ in \eqref{eq:consistent1}; the red and blue area together depict the set $\overline{\Upsilon}_S(\epsilon)$ in \eqref{eq:overExpression}; the green area shows the set $\mathcal{E}_{T}$ in \eqref{eq:consistent2}.}\label{fig:FirstOrder_Dis}
\end{center}	
\end{figure}

The key sets are depicted in Figure \ref{fig:FirstOrder_Dis}. Again we can see that stabilizing all the systems inside the intersection $\mathcal{E}_{T} \cap \overline{\Upsilon}_S(\epsilon)$ is clearly less conservative than stabilizing all systems belonging to $\mathcal{E}_{T}$. Moreover, solving Theorem \ref{thm:sufficientControllerNoise} by CVX, we compute $K = -24.63$, which stabilizes all the system matrices $(A,B)$ contained in the intersection, leading to a Schur stable closed-loop matrix $A_{T \star} + B_{T \star} K$. 

\subsection{High order unstable system}\label{sec:hous}

In this subsection, we consider the source system and the target system to be both the unstable batch reactor processes in \cite[Section 2.6]{Green1995}. The linearized continuous-time model of the actual source system is $\dot{x}_S = G_{S \star} x_S + H_{S \star} u_S$, where 
\begin{equation*} 
G_{S \star} \! = \! \left[\begin{smallmatrix} 1.38 & -0.2077 & 6.715 & -5.676 \\ -0.5814 & -4.29 & 0 & 0.675 \\ 1.067 & 4.273 & -6.654 & 5.893 \\ 0.048 & 4.273 & 1.343 & -2.104 \end{smallmatrix}\right], H_{S \star} \! = \! \left[\begin{smallmatrix} 0 & 0 \\ 5.679 & 0 \\ 1.136 & -3.146 \\ 1.136 &  0 \end{smallmatrix}\right].
\end{equation*}
Assuming that the control input is piecewise constant over the sample time, we can discretize the continuous-time system by the zero-order holder method. The measured state $x_S$ is corrupted with energy bounded disturbances $d_S = [d_{S1}, d_{S2}, d_{S3}, d_{S4}]^{\top}$. The discretized actual source system is then given by
\[x_S(i+1) = A_{S \star} x_S(i) + B_{S \star} u_S(i) + d_S(i), \quad i\in \mathbb{N}_0.\]

\begin{figure}[t]
\begin{center}
    \includegraphics[width=0.8\columnwidth]{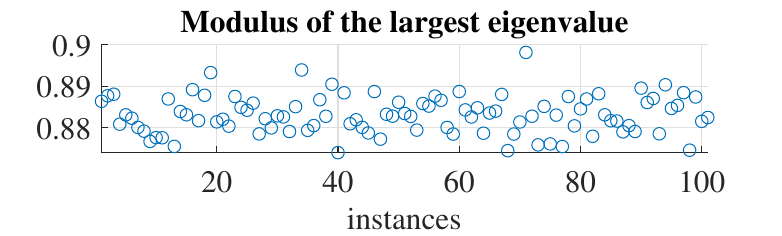}
    \vspace{-0.3cm}
    \caption{High-order systems; the largest eigenvalues of randomly selected systems from \eqref{eq:problem2solve-3}. }\label{fig:HighOrder_DisFree}
\end{center}	
\end{figure}

\begin{figure}[t]
\begin{center}
    \includegraphics[width=0.8\columnwidth]{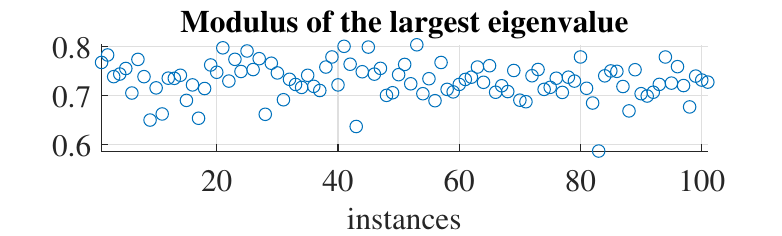}
    \vspace{-0.3cm}
    \caption{High-order systems; the largest eigenvalues of randomly selected systems from \eqref{eq:problem2solveNoise-3}.}\label{fig:HighOrder_Dis}
\end{center}	
\end{figure}

The simulation settings are as follows. The control and sampling period is 0.01s. First, we assume the source system is disturbance free. For the target system, we set $\epsilon = 0.1$, meaning that the actual target system is 0.1-close to the source system. We arbitrarily select one instance of $(A_{T \star}, B_{T \star})$ from $\mathcal{E}_S(+\epsilon)$ and apply the input $u_T \thicksim U[-50, 50]$ and the independent identically distributed disturbances $(d_{T1}, d_{T2}, d_{T3}, d_{T4}) \thicksim U[-0.02,0.02]$. Note that the value of $(A_{T \star}, B_{T \star})$ is only used for generating data, but not for the controller design. The data length is set to $N_T = 4$. Bearing in mind \eqref{eq:D0}, the upper bound on the disturbance (i.e., 0.02) and the data length (i.e., 4), we have $\Delta_{T} \Delta_{T}^{\top} = 0.0064I_4$.

Solving Theorem \ref{thm:sufficientController} by CVX, we have 
\[ K = \left[\begin{smallmatrix} 15.041 & 5.459 & -4.688 & -6.380 \\ 20.782 & 5.918 & 19.304 & 17.215 \end{smallmatrix}\right]. \] 
In order to numerically demonstrate that the controller $K$ indeed works, we select $(A_{T \star}, B_{T \star})$ and other 100 random system matrices $(A,B)$ inside the intersection $\mathcal{E}_{T} \cap \mathcal{E}_S(+\epsilon)$. For each system instance, we plot the corresponding closed-loop eigenvalue with largest modulus in Figure \ref{fig:HighOrder_DisFree}. As can be seen from the figure, all the picked instances of possible target systems are Schur stable as expected. 

Next, we consider the case where the source system is affected by disturbances. We apply the input $u_S \thicksim U[-50, 50]$ and independent identically distributed disturbances $(d_{S1}, d_{S2}, d_{S3}, d_{S4}) \thicksim U[-0.01,0.01]$ to obtain the data from the source system. The data length is $N_S = 100$. Bearing in mind \eqref{eq:D0}, the upper bound on the disturbance (i.e., 0.01) and the data length (i.e., 200), we have $\Delta_{S} \Delta_{S}^{\top} = 0.04I_4$. For the target system, its simulation settings are the same as before.

Solving Theorem \ref{thm:sufficientControllerNoise} by CVX, we have 
\[ K = \left[\begin{smallmatrix} 14.754 & 10.045 & -6.341 & -11.016 \\ 23.117 & 8.314 & 20.293 & 18.505 \end{smallmatrix}\right]. \] 
We choose $(A_{T \star}, B_{T \star})$ along with other 100 random system matrices $(A,B)$ inside the intersection $\mathcal{E}_{T} \cap \overline{\Upsilon}_S(\epsilon)$. For each system instance, we plot the corresponding closed-loop eigenvalue with largest modulus in Figure \ref{fig:HighOrder_Dis}, which demonstrates that all the instances in the intersection are Schur stable.

\subsection{Comparison}

In this subsection, we show that the our approach is far less conservative than a direct application of classic robust stabilization, namely, to stabilize all systems in $\mathcal{E}_S(+\epsilon)$ 
or to directly stabilize all systems in $\mathcal{E}_T$ by a data-driven LMI \cite{Henk2022}.

For the first comparison, consider the case of a disturbance free source system with the same model as in Subsection \ref{sec:hous}. We investigate 100 different values for the closeness metric $\epsilon$; in particular, we set $\epsilon \in$ \texttt{logspace(-2,1,100)}, which is a logarithmically spaced set of points starting from $10^{-2}$ and ending at $10^{1}$ with 100 points. 
For each value of $\epsilon$, we look for a controller that stabilizes the set
of systems $\mathcal{E}_S(+\epsilon)$. This
problem falls in the realm of model-based robust control
as $\mathcal{E}_S(+\epsilon)$
is completely characterized by $A_{S \star}$, $B_{S \star}$ and $\epsilon$. As such,  necessary and sufficient conditions for the solvability of the problem exist, see e.g.~\cite{Carsten2004,PETERSEN1987351}. For each $\epsilon$ we record whether or not these conditions are feasible.

\begin{figure}[ht]
\begin{center}
    \includegraphics[width=0.82\columnwidth]{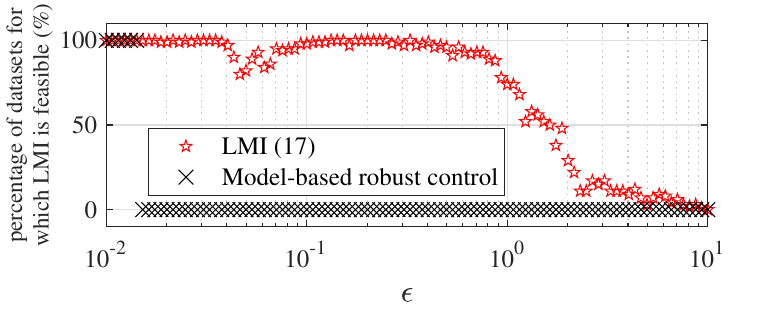}
    \vspace{-0.3cm}
    \caption{Comparison with a model-based robust stabilization.}\label{fig:HighOrder_DisFree_comp_Peter}
\end{center}	
\end{figure}

As for the target system, we also use the same settings as in Subsection \ref{sec:hous}. For each value of $\epsilon$, we generate 100 independent target system's data sets. For each of such data sets, we seek for a controller stabilizing the intersection $\mathcal{E}_T \cap \mathcal{E}_S(+\epsilon)$ via \eqref{eq:sufficientController}, and record whether the LMI is feasible or not. The result is demonstrated in Figure \ref{fig:HighOrder_DisFree_comp_Peter}. As can be seen from the figure, the proposed method substantially reduces conservatism in the design.

\begin{figure}[ht]
\begin{center}
    \includegraphics[width=0.82\columnwidth]{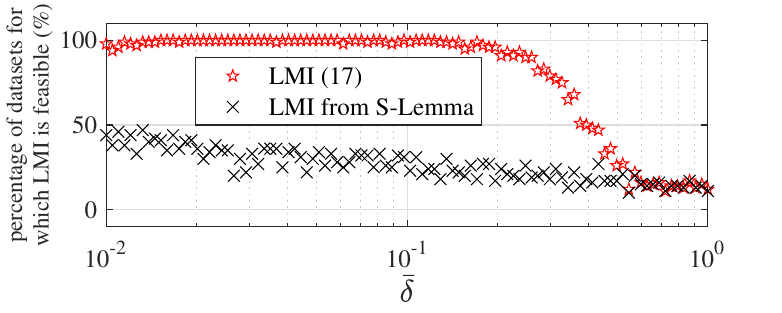}
    \vspace{-0.3cm}
    \caption{Comparison with a direct data-driven robust stabilization of the target system.}\label{fig:HighOrder_DisFree_comp_Slma}
\end{center}	
\end{figure}

For the second comparison, consider the case of a disturbance free source system with the same model as in Subsection \ref{sec:hous}. The same settings in Subsection \ref{sec:hous} apply to the source and target system, including $\epsilon = 0.1$ and $N_T = 4$. The only difference is the noise energy for the target system, where we set $(d_{T1}, d_{T2}, d_{T3}, d_{T4}) \thicksim U[-\bar{\delta},\bar{\delta}]$, with $\bar{\delta} \in$ \texttt{logspace(-2,0,100)}. For each value of $\bar{\delta}$, we generate 100 independent target system's data sets. For each of such data sets, we seek for a controller stabilizing the intersection $\mathcal{E}_T \cap \mathcal{E}_S(+\epsilon)$ via \eqref{eq:sufficientController}, and for a controller stabilizing the set $\mathcal{E}_T$ via \cite{Henk2022}, and record whether the associated LMI is feasible or not. The result is demonstrated in Figure \ref{fig:HighOrder_DisFree_comp_Slma}, which shows the advantage of the proposed method over a direct robust stabilization of the target system by the S-lemma procedure.

\section{Conclusion}\label{sec:Conclusion} 
We have shown how to design a stabilizing controller for a (target) systems without having an accurate model or sufficient data. Instead, our design relies on a second system, termed ``source" for which abundant data is available.
Under suitable assumptions, we were able to transfer the the knowledge learnt from the source system to stabilize the target system. Our solution is much less conservative than stabilizing the entire set of systems consistent with the target system data which is often infeasible. 
We envision this approach to be particularly useful when the target system is an unstable physical system, whereas the source one serves as its digital twin or a simplified lab model. In this case, unlike the source system, collecting sufficient data from the target system is challenging as its state variables will grow exponentially in open-loop experiments.

We stress that only the stabilization problem as a prototypical control problem was considered in this note, and studying other control problems like $\mathcal{H}_2$ and $\mathcal{H}_{\infty}$ control using the proposed transfer learning idea is of interest for future research. We also foresee the potential to extend the proposed idea to certain classes of nonlinear systems.
\appendix

To prove Lemma \ref{lm:Matrix-S-S}, we need a few technical results. We first revisit \cite[Theorem 4.1]{Polyak1998}:
\begin{theorem}\label{thm:Polyak1998} 
Let $g_i : \mathbb{R}^{p} \to \mathbb{R}$, $i = 0,1,2$, be homogeneous quadratic functions, i.e., $g_i(z) = z^{\top} \mathbf{A}_i z$. 
Suppose that $p \geq 3$ and 
\begin{align} 
    & \exists \mu_1, \mu_2 \in \mathbb{R} \ \text{such that} \ \mu_1 \mathbf{A}_{1} + \mu_2 \mathbf{A}_{2} \succ 0, \\
    & \exists \bar{z}\in\mathbb{R}^{p} \ \text{such that} \ g_1(\bar{z}) < 0 , \ g_2(\bar{z}) < 0. \label{eq:T1-2}
\end{align}	
Then, 
\begin{equation}\label{eq:T2-1}
    g_0(z) \leq 0, \quad \forall z \in \mathcal{I}_1 := \{ z : g_i(z) \leq 0, \, i=1,2 \}
\end{equation}
if and only if there exist $\tau_1 \geq 0$, $\tau_2 \geq 0$ such that
\begin{equation}\label{eq:T3-1}
    \mathbf{A}_{0} \preceq \tau_1 \mathbf{A}_{1} + \tau_2 \mathbf{A}_{2}.
\end{equation}
\end{theorem}

\medskip{}
The following lemma discusses a strict version of \eqref{eq:T2-1}, where the points on the boundary of $\mathcal{I}_1$ are excluded from the result.

\begin{lemma}\label{cor:Polyak1998}
Let $g_i : \mathbb{R}^{p} \to \mathbb{R}$, $i = 0,1,2$, be homogeneous quadratic functions, i.e., $g_i(z) = z^{\top} \mathbf{A}_i z$.  
Suppose that $p \geq 3$ and
\begin{align} 
    & \exists \mu_1, \mu_2 \in \mathbb{R} \ \text{such that} \ \mu_1 \mathbf{A}_{1} + \mu_2 \mathbf{A}_{2} \succ 0, \label{eq:C1-1} \\
    & \exists \bar{z}\in\mathbb{R}^{p} \ \text{such that} \ g_1(\bar{z}) < 0 , \ g_2(\bar{z}) < 0. \label{eq:C1-2}
\end{align}	
Then, 
\begin{equation}\label{eq:C2-1}
    g_0(z) \leq 0, \quad \forall z \in \mathcal{I}_2 := \{ z : g_i(z) < 0, \, i=1,2 \}
\end{equation}
if and only if there exist $\tau_1 \geq 0$, $\tau_2 \geq 0$ such that
\begin{equation}\label{eq:C3-1}
    \mathbf{A}_{0} \preceq \tau_1 \mathbf{A}_{1} + \tau_2 \mathbf{A}_{2}.
\end{equation}
\end{lemma}

\begin{pfof} {Lemma \ref{cor:Polyak1998}}
The `if' part trivially follows from Theorem \ref{thm:Polyak1998}. For the `only if' part, our strategy is to prove \eqref{eq:C2-1} $\Rightarrow$ \eqref{eq:T2-1}. To this end, it suffices to show that if \eqref{eq:C2-1} holds, then $g_0(z) \leq 0$, for all $z \in \, \mathcal{I}_{bd}$ where
\begin{align*}
\mathcal{I}_{bd} & := \mathcal{I}_1 \setminus \mathcal{I}_2\\
= & \{ z : g_1(z) \! = \! 0, g_2(z) \! \leq \! 0 \} \cup \{ z : g_1(z) \! \leq \! 0, g_2(z) \! = \! 0 \}.
\end{align*}
We show this claim by using a contradiction argument. 
First, we note that, by \cite[Theorem 2.1]{Polyak1998} and condition \eqref{eq:C1-1}, the set \[\mathcal{G} := \{ [g_0(z) \ g_1(z) \ g_2(z)]^{\top} : z \in \mathbb{R}^{p} \} \subseteq \mathbb{R}^{3} \]
is a {\textit{convex} cone}.
Now, suppose that there exists $z^{+} \in \mathcal{I}_{bd}$ such that $g_0(z^{+}) > 0$. Let $\alpha\in (0, 1)$. Since both $[g_0(z^{+}) \ g_1(z^{+}) \ g_2(z^{+})]^{\top}$ and $[g_0(\bar{z}) \ g_1(\bar{z}) \ g_2(\bar{z})]^{\top}$ are in the convex set $\mathcal{G}$, there exists $\hat z\in \mathbb{R}^{p}$ such that
\[ \alpha \left[ \begin{array}{c} g_0(\bar{z}) \\ g_1(\bar{z}) \\ g_2(\bar{z}) \end{array} \right] + (1-\alpha) \left[ \begin{array}{c} g_0(z^{+}) \\ g_1(z^{+}) \\ g_2(z^{+}) \end{array} \right] = \left[ \begin{array}{c} g_0(\hat{z}) \\ g_1(\hat{z}) \\ g_2(\hat{z}) \end{array} \right] \in \mathcal{G}. \]
Bearing in mind $z^+\in \mathcal{I}_{bd}$, $g_0(z^+)>0$, $\bar z \in \mathcal{I}_2$, $g_0(\bar z)\leq 0$, and \eqref{eq:C1-2}, we have 
$g_0(\hat{z}) > 0$, $g_1(\hat{z}) < 0$ and $g_2(\hat{z}) < 0$ for sufficiently small $\alpha$. The latter contradicts \eqref{eq:C2-1}, and completes the proof. 
\end{pfof}

\medskip{}
Next, the result of Lemma \ref{cor:Polyak1998} is generalized to matrix variables.
\begin{lemma}\label{lm:homogeneousMQ} 
Let $g_i : \mathbb{R}^{p \times q} \to \mathbb{R}^{q \times q}$, $i = 0,1,2$, be homogeneous matrix quadratic functions, $g_i(Z) = Z^{\top} \mathbf{A}_i Z$. Suppose that $p \geq 3$ and 
\vspace{-1mm}
\begin{align} 
    & \exists \mu_1, \mu_2 \in \mathbb{R} \ \text{such that} \ \mu_1 \mathbf{A}_{1} + \mu_2 \mathbf{A}_{2} \succ 0, \label{eq:B1-1} \\
    & \exists \bar{Z}\in\mathbb{R}^{p \times q} \ \text{such that} \ g_1(\bar{Z}) \prec 0,\ g_2(\bar{Z}) \prec 0. \label{eq:B1-2}
\end{align}	
Then, the following statements are equivalent:
\begin{itemize}
    \item[(i)] $g_0(Z) \preceq 0$, $\forall Z : g_i(Z) \preceq 0, \, i=1,2$.
    \item[(ii)] $g_0(Z) \preceq 0$, $\forall Z : g_i(Z) \prec 0, \, i=1,2$. 
    \item[(iii)] $\exists$ $\tau_1 \geq 0$, $\tau_2 \geq 0$ such that $\mathbf{A}_{0} \preceq \tau_1 \mathbf{A}_{1} + \tau_2 \mathbf{A}_{2}$.
\end{itemize}
\end{lemma}

\begin{pfof} {Lemma \ref{lm:homogeneousMQ}}
Clearly, (i) $\Rightarrow$ (ii) and (iii) $\Rightarrow$ (i). Hence, it suffices to prove the implication (ii) $\Rightarrow$ (iii). 
Suppose (ii) holds and let $z \in \mathbb{R}^{p}$ be such that $z^{\top} \mathbf{A}_{i} z < 0, \, i=1,2$. Then, following the footsteps of the proof of \cite[Theorem 7]{Henk2022}, we can show that $z^{\top} \mathbf{A}_{0} z \leq 0$, in other words, \eqref{eq:C2-1} holds. {Note that \eqref{eq:B1-1} is identical to \eqref{eq:C1-1}, and \eqref{eq:B1-2} implies \eqref{eq:C1-2}. } Then we conclude by Lemma \ref{cor:Polyak1998} that condition (iii) is satisfied.
\end{pfof}

\medskip{}
Finally, we use the result of Lemma \ref{lm:homogeneousMQ} to prove Lemma \ref{lm:Matrix-S-S}.

\begin{pfof} {Lemma \ref{lm:Matrix-S-S}}
Obviously, (I) $\Rightarrow$ (II) and (III) $\Rightarrow$ (I). Thus, it suffices to prove the implication (II) $\Rightarrow$ (III). Suppose that (II) holds and let $Z \in \mathbb{R}^{(p+q) \times q}$ be such that $Z^{\top} \mathbf{A}_{Qi} Z \prec 0, \, i=1,2$. Then, we can mimic the arguments in the proof of \cite[Theorem 9]{Henk2022} to  conclude that $Z^{\top} \mathbf{A}_{Q0} Z \preceq 0$. Therefore, statement (ii) of Lemma \ref{lm:homogeneousMQ} is satisfied. This implies that statement (iii) of Lemma \ref{lm:homogeneousMQ} or equivalently the statement (III) of Lemma \ref{lm:Matrix-S-S} holds, which completes the proof of this lemma.
\end{pfof}

\bibliographystyle{IEEEtran}
\bibliography{mybib}

\end{document}